\documentclass[10pt,conference,letterpaper]{IEEEtran}
\usepackage{times}
\usepackage[]{todonotes}
\usepackage[margins]{trackchanges}
\usepackage[cmex10]{amsmath}
\usepackage{graphicx}
\usepackage{array}
\usepackage{mdwtab}
\usepackage{hyperref}
\usepackage{booktabs}
\usepackage{balance}
\usepackage{flushend}

\begin{document}
\pagenumbering{arabic} 
\pagestyle{plain}

\title{On and Off-Blockchain Enforcement Of Smart Contracts}

\author{
\IEEEauthorblockN{Carlos Molina-Jimenez}
\IEEEauthorblockA{Computer Laboratory\\
University of Cambridge, UK\\
Email: carlos.molina@cl.cam.ac.uk}\\   

\IEEEauthorblockN{Irene Ng\\}
\IEEEauthorblockA{Hat Community Foundation\\
Cambridge, UK\\
Email: irene.ng@hatcommunity.org\\}
 
\and
\IEEEauthorblockN{Ellis Solaiman}
\IEEEauthorblockA{School of Computing\\
Newcastle University, UK\\
Email: ellis.solaiman@ncl.ac.uk}\\

\IEEEauthorblockN{Jon Crowcroft\\}
\IEEEauthorblockA{Computer Laboratory\\
University of Cambridge, UK\\
Email: jon.crowcroft@cl.cam.ac.uk\\} 

\and
\IEEEauthorblockN{Ioannis Sfyrakis\\}
\IEEEauthorblockA{School of Computing\\
Newcastle University, UK\\
Email: ioannis.sfyrakis@ncl.ac.uk}

}

\maketitle
\thispagestyle{empty}

\begin{abstract}

In this paper we discuss how conventional business contracts can be converted into smart contracts---their electronic equivalents that can be used to systematically monitor and enforce contractual rights, obligations and prohibitions at run time. 
We explain that emerging blockchain technology is certainly a promising platform for implementing smart contracts but argue that there is a large class of applications,  where blockchain is inadequate due to
performance, scalability, and consistency requirements, and also due to language expressiveness and cost issues that are hard to solve. We explain that in some situations a centralised approach that does not rely on blockchain is a better alternative due to its simplicity, scalability, and performance. 
We suggest that in applications where decentralisation and transparency are essential, developers can advantageously combine the two approaches into hybrid solutions where some operations are enforced by enforcers deployed on--blockchains and the rest by enforcers deployed on trusted third parties.  

\end{abstract}

\section{Introduction}

  This paper focuses on scenarios where two or more parties interact with each other to conduct business over the Internet. Typical scenarios involve consumers and providers where the latter sell tangible items or computing services to the former. A specific example is the selling of personal data collected from IoT sensors or social media applications to data consumers.  

We assume that the business parties involved are reluctant to trust each other unguardedly; that is, without software mechanisms that assure that 1) all parties act in accordance with some agreed upon rules, and 2) performed actions are indelibly recorded on means that make them undeniable and  examinable, for example, to determine the sequence of actions that led to an unexpected outcome and subsequent dispute.

In conventional business, the mechanisms normally used in these situations are business contracts supported by \textit{ledgers}. The contract stipulates what actions the parties are expected to execute, while the ledger is used to record the history of the actions that have been executed. It is widely accepted that equivalent mechanisms are also needed in electronic business. An emerging solution that is currently being explored to address this question is \textbf{smart contracts} built on the basis of blockchain technologies~\cite{Kieron2017}~\cite{Bartoletti2017}. Examples of such technologies are Bitcoin~\cite{AndreasAntonopoulos2017},
Ethereum~\cite{Ethereum2017} and Hyperledger~\cite{HyperledgerHome}. However, blockchain-based smart contracts are only at their initial research stage, and plagued with questions about their scalability, performance, transaction costs and other questions that emerge from their descentralised nature.
 
This article makes the following contributions to help clarify some of these issues. 
i) We explain that there are different approaches to implement smart contracts ranging from centralised to decentralised. ii) We explain the advantages and disadvantages of these approaches and argue that their suitability in solving the 
problem depends on the particularities of the application, the assumptions made about 
the application, and the facilities offered by the blockchain technology available. iii) We 
argue that there is a large class of applications that can benefit from a hybrid solution.
 
The remainder of this article is organised as follows: Section~\ref{motivatingscenario} presents
a contract example to motivate the use of smart contracts. In Section~\ref{background}, we
introduce smart contracts and describe the difference between the centralised and decentralised variations. Section~\ref{implementations}
discusses implementation alternatives of smart contracts (the main contribution of the paper). 
Section~\ref{relatedwork} places our work within past and current contexts. 
In Section ~\ref{conclusions}, we present some concluding remarks and raise questions that
in our view, need research attention.

\section{Motivating scenario}
\label{motivatingscenario}
\begin{figure}[!t]
	\centering
	\includegraphics[width=0.85\columnwidth]{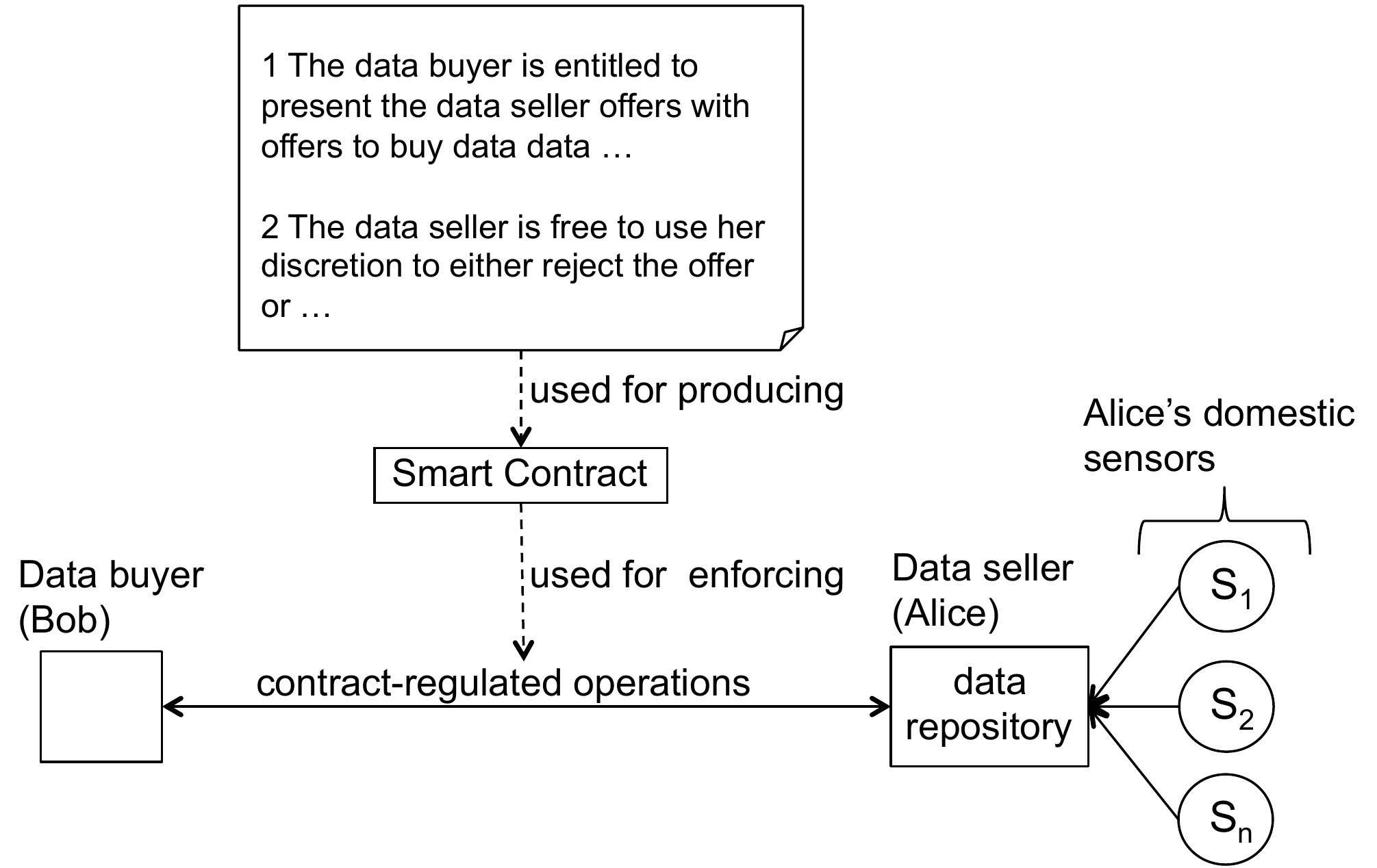}
	\caption{Data trading regulated by a smart contract.}
	\label{fig:datatradingscenario}
\end{figure}

An illustrative example of a contractually regulated IoT application of our research 
interest is shown in Fig.~\ref{fig:datatradingscenario}.

Alice is a person in possession of personal data that she would like to sell and as such
she plays the role of a \emph{Data seller}. The \emph{Data Buyer} (represented by Bob) is a company interested in buying data from Alice. Alice gathers her data from different sources, such as her social network activities, body sensors and domestic sensors, as envisioned in~\cite{HATrumpelPlatform}.  For simplicity and to frame the discussion, we assume that Alice is trading only her data collected from her domestic sensors.
Like in~\cite{HATrumpelPlatform}, we assume that Alice stores her data in a personal repository, perhaps located in the cloud. Like in the "Hat" project~\cite{HATHome}, we assume that Alice is the absolute owner of the data and that she is entitled to negotiate with potential data buyers how to trade her data, i.e., to whom to sell it to, when, and under which conditions. The negotiation process can be as sophisticated as needed. Since this issue falls outside of the scope of this paper, we consider only a simple \emph{accept or reject the offer as it is} negotiation process. 

As explained in~\cite{Kieron2017}, realistic conventional legal contracts are complex documents, written for example in English. 
Normally these documents include inconsistencies and ambiguities that are tolerable because they are expected to be interpreted with the help of human judgment. However because contracts do contain inconsistencies and ambiguities, their full conversion to electronic equivalents is 
a challenging task that falls outside the ambitions of this paper. However we refer the reader to previous research efforts in this direction~\cite{solaiman2003}~\cite{Solaiman2015}~\cite{SolaimanSun2015}~\cite{Solaiman2016}. The focus of our work
is on specific clauses of the contracts that are stipulated sufficiently precise that makes them 
amenable to computer language encoding. 

We believe that to be of practical use, a smart contract needs to include clauses
that take into consideration normal and undesirable paths of the business process. The latter 
account for the occurrence of exceptional situations. Examples of exceptions in 
our example are failures to deliver the payment or the data before a deadline 
or failure to deliver a valid payment or data of the expected quality. An example of contractual clauses that Alice and Bob can use to regulate their
data trading are the following:

\begin{it}
\begin{enumerate}
\item The buyer (\underline{Bob}) is entitled to present the data seller (\underline{Alice})
       with \textbf{offers to buy data} collected from Alice's domestic sensors.

\item The data seller is free to use her discretion to either \textbf{reject the offer} or 
      \textbf{accept the offer} as it is.
 \begin{enumerate}
   \item The data seller 
         is expected to \textbf{send a notification of
         offer acceptance} within 36 hrs of receiving 
         the offer, when she decides to accept it.
    \item Failure to send a notification will be
          considered as offer rejection.
  \end{enumerate}

\item The data buyer is obliged to \textbf{send the payment} to the data seller within 24 hrs of 
       receiving the notification of acceptance. 
\begin{enumerate}
 \item Failure to meet his obligation will result in an 
       abnormal termination of the agreement to be sorted 
       out off line.
\end{enumerate}

\item The data seller is obliged to \textbf{send a notification of
      payment acceptance} to the data buyer within 24 hrs of 
      collecting the payment.
      \begin{enumerate}
       \item Failure to meet his obligation will result in an 
       abnormal termination of the agreement to be sorted 
       out off line.
       \end{enumerate}

\item The data seller is obliged to \textbf{make the data available} to the data 
       seller within 24 hrs of collecting the payment and \textbf{maintain the data 
       repository accessible} during the following seven days.
 \item The Data buyer is entitled to \textbf{place data requests} 
        against the data seller repository without exceeding 24 
        data requests per day.
  \item The data buyer is entitled to \textbf{close the repository}
        upon expiration of the seven day period.
        
\item This agreement will be considered successfully complete
      when the seven day period expires.

\end{enumerate}
\end{it}
 
 The clauses include several contractual operations that we have highlighted in bold  such as \emph{offer to buy data}, \emph{reject the offer}, \emph{accept the offer}, \emph{send a notification of offer acceptance}, \emph{send payment}, etc. Though the clauses are relatively simple, they are realistic enough to illustrate our arguments.

\section{Smart contracts: background}
\label{background}

A smart contract is an event--condition--action stateful computer program, executed between two or more parties that are reluctant to trust each other unguardedly. It can be regarded as Finite State Machine (FSM) that keeps a state that models the development (from initiation to completion) of a shared activity\cite{Molina03}. For instance, in~\cite{MolinaTSC2011}~\cite{Solaiman20162}, the state is used for modeling changes in rights, obligations and prohibitions as they are fulfilled or violated by the parties.

Research on executable contracts can be traced back to the mid
80s and early 90s~\cite{Minsky1985,Lindsay1993}. In 1997,
Szabo used the term smart contract~\cite{Szabo1997} to refer
to contracts that can be converted into computer code and
executed. However, commercial interest in smart contracts
emerged only in 2008 motivated by the publication of Satoshi's 
Bitcoin paper~\cite{Satoshi2008} that inspired the 
development of cryptocurrencies, smart contracts and
other distributed applications. Satoshi departed from the 
centralised approach taken in previous research and demonstrated how
smart contracts can be decentralised.

\subsection{Centralised and decentralised smart contracts}
\begin{figure}[!t]
	\centering
	\includegraphics[width=0.85\columnwidth]{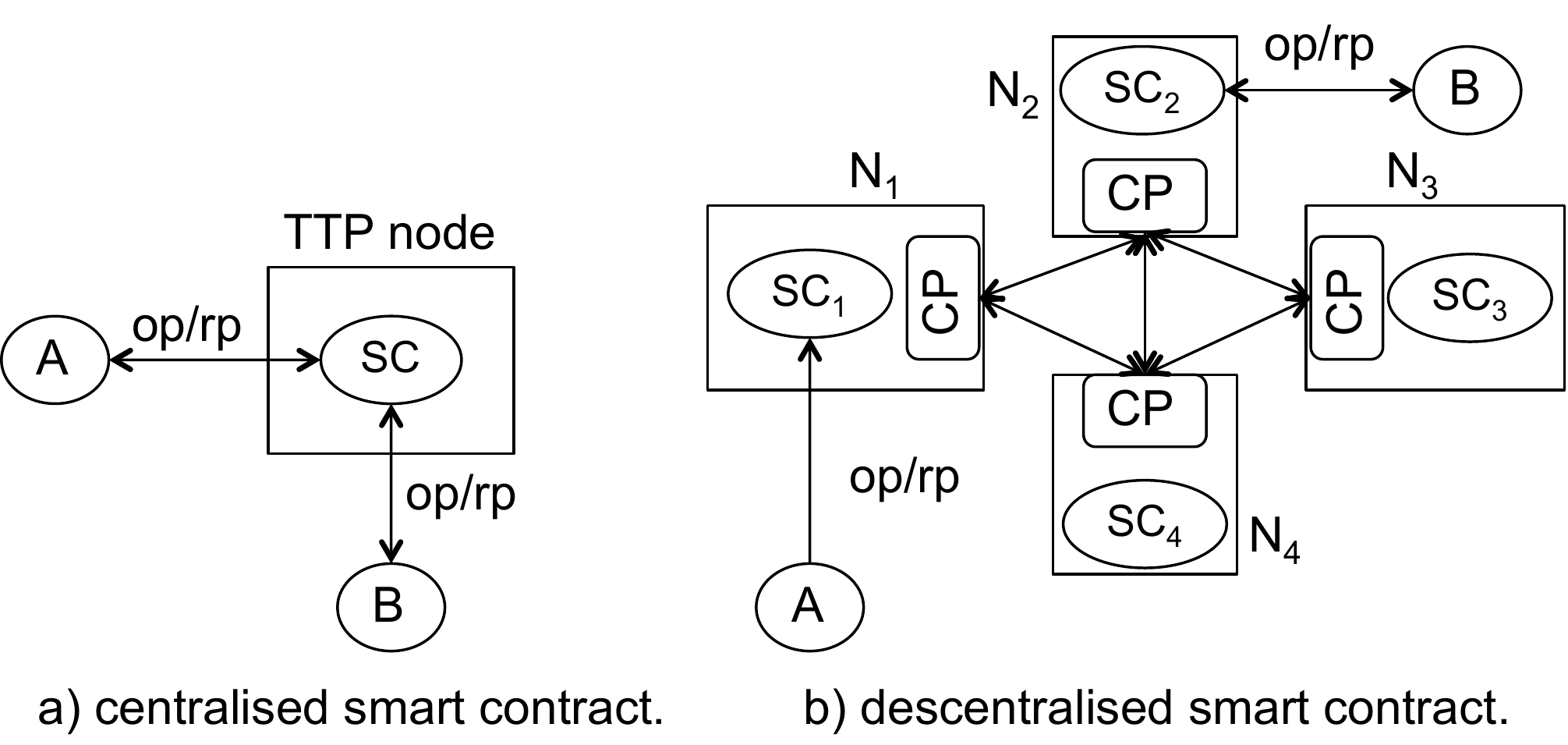}
	\caption{Centralised and descentralised implementation of a smart contract.}
	\label{fig:centralisedanddescentralisedcontract}
\end{figure}

Depending on the number of instances (copies) of the smart contract deployed to monitor and enforce the contract we distinguish between centralised and decentralised (distributed) approaches (Fig.~\ref{fig:centralisedanddescentralisedcontract}). 

In the figure, \emph{A} and \emph{B} are business partners, for example, Alice and Bob of our contract example of Section~\ref{motivatingscenario}. \emph{SC} is the corresponding smart contract. 
 \emph{op} stands for operation executed against \emph{SC}, \emph{rp} is the corresponding response. \emph{TTP node} is a node under the control of a Trusted Third Party. $N_1,\dots,N_4$ are untrusted nodes. \emph{CP} stands for Consensus Protocol. 
As shown in Fig.~\ref{fig:centralisedanddescentralisedcontract}--a), a contract 
can be implemented as a centralised application that uses a single 
instance of the smart contract (\emph{SC}) running in the TTP node. Besides the disadvantages that a TTP introduces (single point of failure, 
trust placed on the TTP, etc.) this approach is comparatively simpler that the 
decentralised approach. The decentralised approach relies on a set of 
untrusted nodes instead of a single TTP that are used for running several identical instances (shown as ${SC}_1,\dots,{SC}_4$) of the smart contract. In this approach, \emph{A} and \emph{B} are free to place their operation against any of the instances. The price that the decentralised approach pays for getting rid of the TTP is that the untrusted nodes must run a consensus protocol to verify that a given operation has been executed correctly, and to keep the 
states of ${SC}_1,\dots,{SC}_4$ identical. Depending on the protocol used, its computational, communication and performance
degradation cost might be unbearable~\cite{Marko2015} or its consistency guarantees inadequate~\cite{PeterBailis2013} to the extent of rendering 
the decentralised approach unsuitable.

\section{Implementation alternatives}
\label{implementations}
We will take the example of Section~\ref{motivatingscenario} and highlight the advantages and disadvantages of three implementation alternatives.

\subsection{Centralised implementation}
\label{centralisedcontract}
\begin{figure}[!t]
	\centering
	\includegraphics[width=0.85\columnwidth]{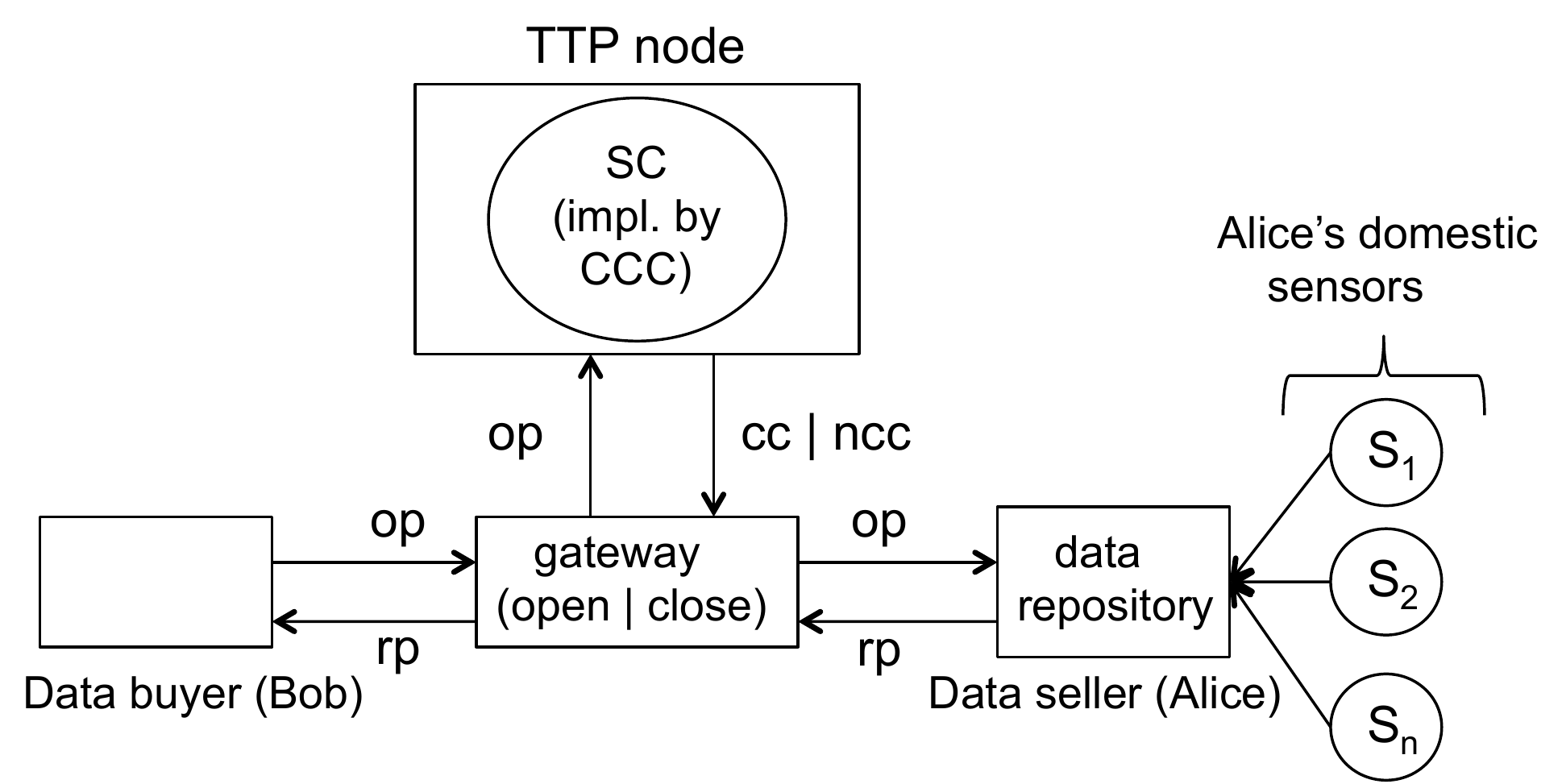}
	\caption{Centralised smart contract.}
	\label{fig:centralisedcontract}
\end{figure}

A centralised implementation is shown in Fig.~\ref{fig:centralisedcontract}.
The role of the \emph{SC} is played by the CCC (Contract Compliance Checker) developed at the University of Newcastle. We use CCC and SC  synonymously in this section. The CCC is a FSM written in Java that accepts 
contractual clauses encoded as business rules written in the Drools 
language~\cite{MolinaTSC2011}. The state of the FSM is altered by
the execution of contractual operations (\emph{op}) initiated by
the business partners, such as \emph{offer to buy data},
and \emph{send the payment}. 
The FSM running within the CCC keeps track of the state  of the business 
process executed between Bob and Alice, and on this
basis it determines if a given operation is contract compliant
(\emph{cc}) or non contract compliant (\emph{ncc}). 
The CCC is used  to control the \emph{gateway}
that grants access to Alice's data. For example, when Bob wishes to
access Alice's data, he i) issues the corresponding operation 
against the gateway, ii) the gateway forwards the operation to the
CCC, iii) the CCC evaluates the operation in
accordance with its business rules that encode the contractual 
clauses and responds with either \emph{cc} or \emph{ncc} to open or 
close the gateway, respectively, iv) the opening of the gateway allows 
Bob's operation to reach the data repository and retrieve the 
response (\emph{rp}) that travels to Bob. Note that, to keep the 
figure simple, the 
arrows show only the direction followed by operations
initiated by Bob. 

It is worth elaborating on the following points. Observe that in the architecture
all the operations are presented to the \emph{SC} for evaluation. The operation
rate is not a problem because the architecture involves only a single instance
of the \emph{SC}, i.e., there is no need to run consensus protocols. Likewise,
the contract clauses are encoded in the Drools languages which are executed
by a FSM implemented in Java. This means that we have a Turing complete programming 
environment that allows us to encode and implement clauses of arbitrary complexity. 
Unfortunately, the centralised approach introduces several drawbacks. For 
example, the contracting parties need to trust the TTP to collect 
undeniable and indelible records of the actions executed by the contracting 
parties and make them available upon request to parties that are entitled to
see them, say to sort out disputes. At the technical level, the TTP node is a single point of failure. Another issue is that the execution of the payment operation is also centralised, we assume a conventional bank card 
payment mediated by a bank as opposed to cryptocurrency payment.

 \subsection{Descentralised implementation}
\label{descentralisedcontract}
\begin{figure}[!t]
	\centering
	\includegraphics[width=0.85\columnwidth]{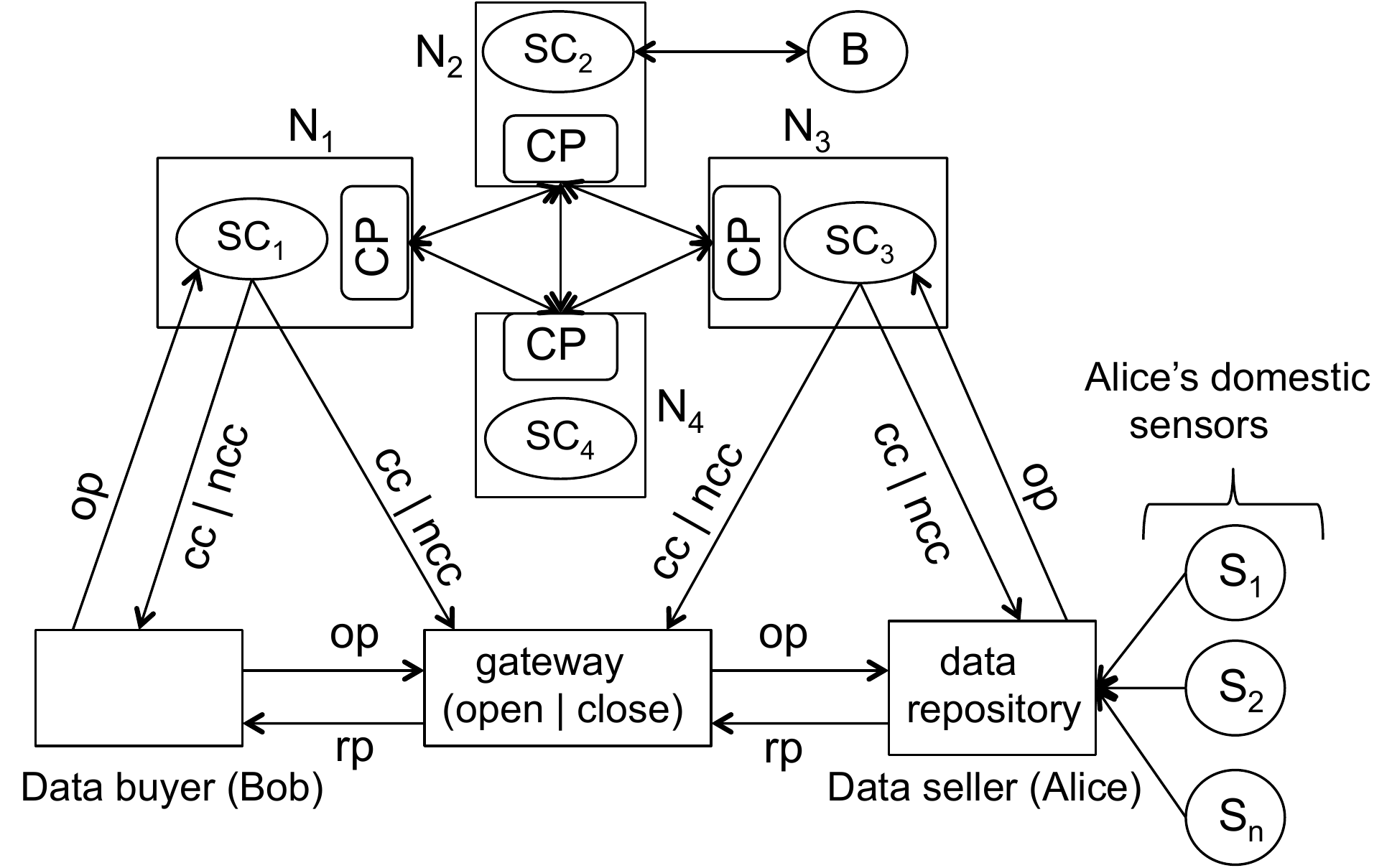}
	\caption{Descentralised smart contract.}
	\label{fig:descentralisedcontract}
\end{figure}

A descentralised architecture  is shown
in Fig.~\ref{fig:descentralisedcontract}. Four instances of the
smart contract (${SC}_1,\dots,{SC}_4$) are deployed in four nodes
$N_1,\dots,N_4$ (one each) of a blockchain platform.

Each operation initiated by a business partner is executed
against the contract; the contract determines if the
operation is contract compliant (\emph{cc}) or non contract 
compliant (\emph{ncc}) and responds to both business partners
accordingly. The response is also sent to the \emph{gateway} 
to open or close it, accordingly.

 To keep the figure simple, we show only the communication lines between 
 the \emph{Data buyer}, ${SC}_1$ and the \emph{gateway}; and between 
 the \emph{Data seller}, ${SC}_3$ and the \emph{gateway}. Yet
 we assume that a given operation can be presented to any of the
 four instances of the smart contract and that any of them can respond
 to the business partners and the \emph{gateway}

The salient feature of the decentralised implementation is the
replication of the smart contract, consequently,
there is no dependency on a single party. The cost to
pay for this benefit is the execution of the consensus
protocol among the instances which can significantly
impact the performance of the smart contract in terms
of number of operations (called transactions in
blockchain terminology) per second that it can analyse,
and the response time to complete a transaction. For example,
Bitcoin, a public blockchain that uses a Proof of Work 
(PoW) consensus algorithm, can only process about 7 
transactions per second. Another problem with Bitcoin 
is its consistency latency: its PoW algorithm offers only 
eventual consistency that might take Bitcoin about
an hour (or longer) to approve and indelibly include a 
transaction in its blockchain~\cite{Decker2016}. Ethereum operating 
under PoW consensus suffer from similar drawbacks. Permissioned 
blockchains like Hyperledger rely on lighter consensus algorithms
such as Proof of State (PoS). However, applications where
eventual consistency is unsafe, demand strong 
consistency~\cite{PeterBailis2013}. Strong consistency can  
only be delivered by communication intensive consensus protocols 
such as Byzantine Fault Tolerant protocols, unfortunately,
these protocols suffer from scalability issues~\cite{Marko2015}. 
Some smart contract applications (for example, applications 
that require instantaneous payment or the delivery of real time
data) fall within this category. Another issue that impacts
decentralised approaches that rely on public blockchains
is the transaction fee incurred by each operation 
analysed by the smart contract. In this order, it would be
insensible to take a decentralised implementation
approach for the contract example of Section~\ref{motivatingscenario}
if the  data buyer was to place a large number of
operations to retrieve small pieces of data under
stringent time constraints.

 \subsection{Hybrid implementation}
\begin{figure}[!t]
	\centering
	\includegraphics[width=0.85\columnwidth]{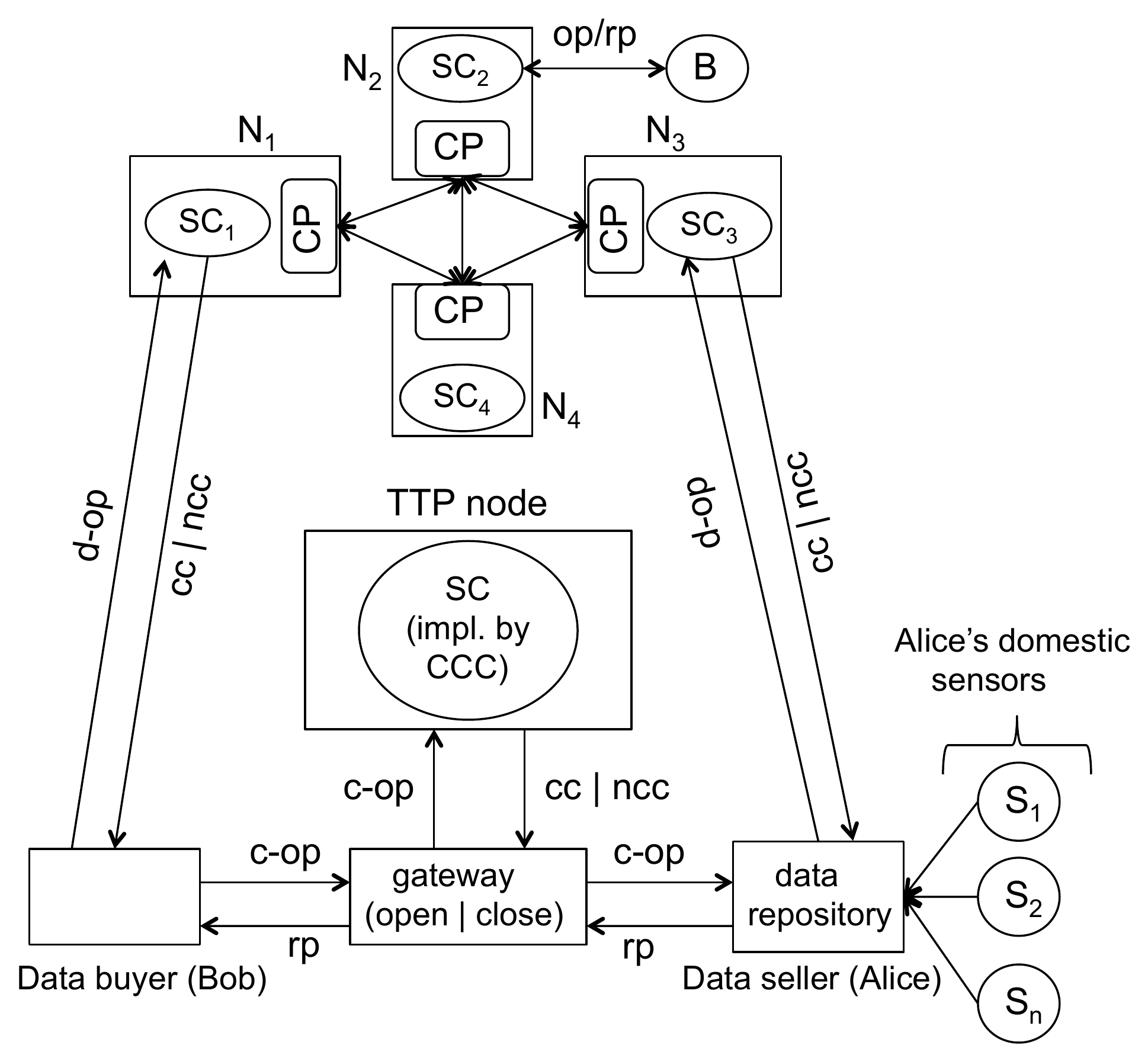}
	\caption{Hybrid smart contract.}
	\label{fig:hybridcontract}
\end{figure}

Fig.~\ref{fig:hybridcontract} shows the architecture of a hybrid
implementation. 

It combines features from 
the centralised and decentralised approaches discussed, respectively, 
in Sections~\ref{centralisedcontract} and 
\ref{descentralisedcontract}. We separate
the contractual operations into two classes: decentralised
operations (\emph{d--op}) that need blockchain support and 
operations that can be executed in a centralised fashion 
(\emph{c--op}). \emph{d--op} operations are encoded
using the descentralised approach and enforced by the 
instances (${SC}_1,\dots,{SC}_4$) whereas operation of the 
\emph{c--op} category are encoded using the centralised approach 
and enforced by the CCC.

The designer separates the contractual
operation into \emph{d--op} and \emph{c--op} on the
basis of several criteria. As examples, we can mention
some key parameters related to the blockchain technology. The
list is meant to be illustrative rather then exhaustive. 
Complementary advise is discussed in~\cite{Eberhardt2017,Guy2015} where
they take into account privacy concerns along with computation and 
data storage costs.
 
One decision criterion is the expressiveness of the language
used for writing the contract. For instance, if the 
blockchain does not offer a Turing--complete language,
the implementers needs to keep the \emph{d--op}
category simple. Bitcoin for example, offers only a
stack--based opcode scripting language that does not support
loops or flow control structures. In contrast, in a
blockchain platform like Ethereum that offers a 
Turing--complete language the designer can afford to
pass as much complexity to the decentralised part
of the figure as she wishes to.
Another decision criterion is the transaction fee which
is an issue in private blockchains like Bitcoin and
Ethereum but not in Hyperledger~\cite{Dominic2014} when it is operated
as a permissioned blockchain. For example,
Bitcoin and Ethereum have already experienced average transaction
fees of 54.90 and 4.15 USD, respectively~\cite{bitinfocharts2018}.
Another central parameter to
take into account is the performance of the
blockchain, for example, the number of transactions per 
second and consistency requirements as explained in 
Section~\ref{descentralisedcontract}. Operations that 
demand strong consistency would be good candidates
to be implemented as \emph{c--op}. The performance of the blockchain is especially relevant to IoT applications where transactions must be automatically monitored to ensure that they perform under strict Quality of Service requirements. For example one could easily imagine an additional clause being added to the contract in Section ~\ref{motivatingscenario} requiring the repository to process each request for data at a particular rate that would be too fast to be monitored using a smart contract deployed on a blockchain. In such a scenario, a centralised smart contract would be more logical, whereas the blockchain would be used to record important milestone events such as the sending and receipt of payments for received data. 

We envision that the centralised and descentralised integration
can be operated in several ways, including the following:

\subsubsection{Indelible blockchain--based log}
\label{indelible}
We can operate the blockchain--based
part of Fig.~\ref{fig:hybridcontract} as a passive 
log that records events that the
parties consider worth duplicating in the
blockchain as well as in the TTP node. By
passive we mean that  ${SC}_1,\dots,{SC}_4$ are
not involved in enforcing activities---this
is entirely the responsibility of the CCC. This
arrangement is useful when one or more of
the contracting parties is reluctant to 
trust the TTP blindly, say because it is
deployed within the buyer's premises---
currently a common business practice~\cite{MolinaSOCA2011}.
In this arrangement, the \emph{d--op} set will 
include operations aimed at creating additional
records while \emph{c--op} will include all
the contractual operation like in~\ref{centralisedcontract}.
The CCC and ${SC}_1,\dots,{SC}_4$ operate independently
from each other.

\subsubsection{Cryptocurrency--based payment channel}
\label{cryptocurrencypay}
The data buyer of the example of Section~\ref{motivatingscenario}  
can take advantage of payment services offered by a public blockchain (for example,
Bitcoin) and use the top part of Fig.~\ref{fig:hybridcontract}
to pay in satoshis. This approach is recommended only when the 
payment operation is significantly larger than the transaction 
fees and is not repetitive. In this arrangement,
the \emph{d--op} set will include only the \emph{send the payment}
operation stipulated in clause 3. In this arrangement, the
CCC requires the assistance of the smart contract
running in the blockchain (${SC}_1,\dots,{SC}_4$) only
to verify that the data buyer has fulfilled his obligation
to pay. For instance, the data buyer application can submit
his payment through Bitcoin, wait for the confirmation of his
transaction, collect the evidence and submit it to the
CCC.

\subsubsection{Off-blockchain execution of operations}
\label{offblockchain}
In this arrangement the CCC running in the TTP node
is used as an off the blockchain channel. The designer places in
the \emph{d--op} set only the contractual operations that need
decentralised treatment and leaves the remaining in the 
 \emph{c--op}. Naturally, operations that cannot be executed
 in the decentralised blockchain because of the issues
 discussed in Section~\ref{descentralisedcontract}
 need to be included in \emph{c--op} set. A good candidate operation to
 place in the \emph{d--op} set is \emph{send the payment} (see 
 Section~\ref{cryptocurrencypay}). Another candidate is \emph{close the repository} when the data seller wishes to generate indelible records about
 the closing time of her repository and completion of the contract.
 The remaining operations can be cheaply and efficiently enforced
 by the CCC, the inclusion of \emph{place data requests} (clause 6),
 in the \emph{c--op} set is highly desirable because its recurrence would
 incur high accumulative transaction fees.

It is worth clarifying that there are some similarities between
the deployment shown in Fig.\ref{fig:hybridcontract} and the 
lightning channels for executing off--blockchain payments in 
Bitcoin~\cite{Joseph2016}. However, observe that in lighting
networks the aim is to create channels for conducting
micro--payment operations off the 
blockchain to save on transaction fees. In contrast, 
in Fig.~\ref{fig:hybridcontract} we use the CCC (a complete
contractual enforcing tool) to execute most of the contractual 
operations off--blockchain. Operations from 
both sets are independently converted to smart contracts and
enforced at run time.

\section{Related work}
\label{relatedwork}
Research on smart contracts was pioneered by Minsky in the
mid 80s\cite{Minsky1985} and followed by
Marshall~\cite{Lindsay1993}. Though some of the contract tools
exhibit some descentralised features~\cite{Minsky2010}, those
systems took mainly centralised approaches. Within this category
falls~\cite{GovernatoriEDOC2006} and
\cite{perringodart}. To the same category belongs the model for 
enforcing contractual agreements 
suggested by 
IBM~\cite{ludwig2003soa} and the Heimdhal 
engine~\cite{PedroGama2006} aimed at monitoring state obligations (see clause 5 of the contract example, 
\emph{maintain the data repository accessible}).
Directly related to our work is the Contract Compliant Checker
reported in~\cite{MolinaTSC2011}~\cite{Solaiman20162} which also took a centralised
approach to gain in simplicity at the expense of suffering from
all the drawbacks that TTPs inevitably introduce. Smart contracts were known as executable contracts or electronic contracts in ~\cite{Molina03}~\cite{solaiman2003}~\cite{Molina04}, where the important issues of smart contract representation and verification were discussed. A pioneering
implementation of a descentralised contract enforcer is discussed
in~\cite{Santosh2005}. The central idea of the authors is
the use of a distributed  middleware (one piece associated to each 
party) that is responsible for keeping indelible records
of the operations executed by each party. The
middleware (called a Business to Business object~\cite{NickCook2004})
is in essence an indelible ledger similar in funcionality to
the hyperledger used by current blockchains.

The publication
of the Bitcoin paper~\cite{Satoshi2008} motivated the development
of several platforms for supporting the implementation of decentralised
smart contracts. Platforms in ~\cite{AndreasAntonopoulos2017}, \cite{Ethereum2017}
\cite{HyperledgerHome} and \cite{SergueiPopov2017} are some of the
most representative. A good summary of the features offered
by these and other platforms can be found in~\cite{Bartoletti2017}. 
Though they differ on language expression power, fees and other
features discussed in Section~\ref{descentralisedcontract} they 
are convenient for implementing descentralised smart contracts. 
The hybrid approach that we 
suggest addresses problems that neither the centralised
or descentralised approach can address separately and
was inspired by the off--blockchain payment 
channel discussed in~\cite{Joseph2016,AndreasAntonopoulos2017}.
The concept of logic--based smart contracts discussed in~\cite{Florian2016}
has some similarities with our hybrid approach. They suggest
the use of logic--based languages in the implementation of
 smart contracts capable of performing on--chain and
 off--chain inference. The difficulty with this approach is
 lack of support of logic--based languages in current blockchain technologies.
 In our work, we rely on the native languages offered
 by the blockchain platforms, for example, Ethereum's Solidity.

\section{Conclusions and Future Research Directions}
\label{conclusions}
The central aim of this paper is to argue that conventional business contracts
can be automated (at least partially) and that depending on several factors, 
the centralised approach suits some applications but others demand descentralised
implementations or even hybrid implementations. We are only starting to explore
hybrid implementation of smart contracts, yet on the basis of the study of the
APIs (JSON--RPC) that Bitcoin, Ethereum and Hyperledger offer, 
the idea seems implementable. Also, it is of practical interest as
it would offer a pragmatic answer to the scalability problems that afflict current
blockchain platforms. Equally importantly, this approach opens several research
questions.

Another issue  is the interaction
between the centralised (CCC) and descentalised components. In 
Fig.~\ref{fig:hybridcontract} they cannot communicate directly. We are currently 
working on a version of the CCC that can be deployed as a micro--service 
capable of interacting with the JSON-RPC Client API that blockchain 
technologies offer. Precisely, we are investigating how the hybrid architecture 
can be realised using the Ethereum blockchain and a CCC implemented as a 
decentralised application (DApp)~\cite{dapp2018}.
The relationship (directly or indirectly) between the CCC and the blockchain
raises several questions that need further investigation. They can interact
directly, indirectly, tightly or loosely. Fig.~\ref{fig:hybridcontract} suggests
the latter where, for example, the CCC can fail and recover while 
the \emph{send the payment}
operation is taking place through the block--chain based smart contract (recall
in Bitcoin it might take longer that 24 hrs to complete a transaction). However,
in some applications a tight relationship might be desirable to hold or divert 
the progress of one of the contracts when its counterpart experiences an
exception or fails. The point is about understanding how to separate the
contractual operations into \emph{c--op} and \emph{d--op} in a manner that 
the two contracts collaborate instead of conflicting with each other. For contracts
with scores of clauses, this issue might require the assistance of model--checking
tools to ensure that the whole contractual clauses are consistent and that
the two sets do not conflict with each other~\cite{Abdelsadiq2010,Ilya2018}.

Another issue is the language for writing the contract. It is 
arguably accepted that declarative languages (rule based
languages in particular) are more convenient
than imperative to encode contractual clauses. This
feature is enjoyed by the CCC. However, current blockchain
platforms support only imperative languages (for example
Ethereum's Solidity). This means that in our hybrid approach the 
contract will be written in two different languages which
will make their interaction less intuitive. Ideally blockchain platforms should support declarative languages, or alternatively developers
should be offered a declarative language that can be automatically translated to languages like Solidity or Drools as needed.

\section*{Acknowledgements}
Carlos Molina-Jimenez is currently collaborating with the HAT Community 
Foundation under the support of Grant RG90413 NRAG/536. 
Ioannis Sfykaris was partly supported by the EU Horizon 2020 project 
PrismaCloud (https://prismacloud.eu) under GA No. 644962.

\bibliographystyle{IEEEtran}
\bibliography{./biblio/references.bib}

\end{document}